\documentclass[%
prd%
 ,twocolumn%
 ,secnumarabic%
,amssymb, amsmath,nobibnotes, aps, prl]{revtex4}
\usepackage{color}
\usepackage{graphicx}
\usepackage{multirow}

\begin{document}
\title{Parametric Oscillatory Instability in a Fabry-Perot Cavity of the Einstein Telescope with different mirror's materials}
\author{S.E. Strigin}
\affiliation{M.V. Lomonosov Moscow State University, Faculty of Physics, Moscow 119991, Russia}
\email{strigin@phys.msu.ru}
\begin{abstract}

We discuss the parametric oscillatory instability in a Fabry-Perot cavity of the Einstein Telescope. 
Unstable combinations of elastic and optical modes for two possible configurations of gravitational wave third-generation detector are deduced. The results are compared with the results for gravitational wave interferometers  LIGO and LIGO Voyager.

\end{abstract}

\maketitle

\section{Introduction}

At present, there are some gravitational wave detectors in the world  like LIGO, VIRGO, GEO and others. At the same time, third-generation  gravitational wave detectors are under construction  which have
their advantages and disadvantages.  Such detectors like  Einstein Telescope(ET) are developed to be able to solve more ambitious goals compared to the so-called second-generation gravitational wave detectors. ET will have a sensitivity of the order of magnitude more than the sensitivity of gravitational wave detectors of the second generation. The details of this project 
are still under development, but the first data  have already been marked in \cite{et1,et2,et3}. In this Letter, we consider two possible configurations of the ET with the Fabry-Perot mirrors made of either silicon and sapphire correspondingly.     

It should be noted that the power $W = 3MW$ which circulates in the Fabry-Perot cavities of the ET will help to increase the sensitivity of the detector. In turn, some main parameters of ET are different in comparison with LIGO such as: distance L between the Fabry-Perot cavity mirrors is 10 km (compared to 4 km in LIGO);  beam radius at the surfaces of the mirrors $w$ is 12 cm (compared
to 6 cm in LIGO); the mirror radius is $R = 30cm$(2.5 larger than the laser beam radius on the mirrors). It makes optical pump mode losses due to the diffraction to be very small. The ET is planned to operate at low temperatures, which will reduce the influence of mechanical noises of various kinds (thermal noise in different parts of detectors, etc.) on
the gravitational wave registration.  It also  planned to place the third-generation gravitational wave detector in a horizontal tunnel with two 10km arms about 800 m underground to reduce the level of the seismic sensitivity(see all other parameters of ET  in Table \ref{tab:addlabel}). For simplicity, we use the Gaussian main pump mode profile.

In turn, it is well-known that large values of circulating power W with low mechanical losses in mirrors may
lead to the  effect of parametric oscillatory instability(PI), which was predicted in 2001 by Braginsky et al.\cite{bsv}. 
This effect produces  the excitation  both of optical Stokes high mode with frequency $\omega_1$ and elastic
mirror mode with  frequency $\omega_m$. The energy is taken from   the main optical mode with frequency $\omega_0$, and the sensitivity of the detector decreases. PI effect is similar to the well-known effect of light Mandelstam-Brillouin scattering. It is worth noting that the instability takes place  when the condition $\omega_0\simeq \omega_1+\omega_m$ is fulfilled. The effect of PI in the Fabry-Perot cavities in gravitational wave antennae like ET  starts due to the high optical power of up to $W\simeq 3$MW circulating inside the arm cavities. 

The presence of anti-Stokes optical modes  reduces the effect of PI but can not completely suppress it\cite{ak,bsv2}. 
Parametric oscillatory  is a serious problem for gravitational wave detectors, and thus
you must have information about the combinations of the optical Stokes and elastic modes that may be candidates for PI\cite{gras3,2007,gras1,gras2,gras4,gras5,gras6,2007_1}. The first observation of PI on the Advanced LIGO detectors has been made in 2015\cite{PIobs}, thereby fully confirming the prediction of Braginsky et al.\cite{bsv}. 

It is also well-known that the Stokes modes can be easily calculated analytically for Gaussian optical modes. On the other hand, elastic
modes in the end test masses  of the cylindrical mirrors can be calculated using various numerical packages(further in the calculations we use the COMSOL$^\circledR$). In our analysis we propose that elastic modes of only end mirror are taken into account in Fabry-Perot cavity.
 \begin{table}[t]
  \centering
    \begin{tabular}{ccc}
    \  Parameters  & Silicon & Sapphire   \\
    \hline
    Pump beam profile &  Gaussian & Gaussian  \\
    $\lambda$, nm &1550 & 1064\\
   $L, $m & 10000 & 10000\\
   $R_1,$m &5070 & 5070\\
   $R_2,$m & 5070 & 5070\\
   $w,$mm & 120 & 120\\
   $W,$MW & 3 & 3\\
   $m,$kg & 230 & 410\\
   $T,$K &123 & 123 \\
   $R,$cm &30 & 30\\
   $E,$Pa & $1.64\times  10^{11}$ & $4\times 10^{11}$\\
   $\rho$,$kg/m^3$ & 2300 &4000\\
   $\sigma$ &0.266 & 0.22\\
    \hline
    \end{tabular}%
    \caption{Parameters of the ET's Fabry-Perot
cavity.}
  \label{tab:addlabel}%
\end{table}%
 \section{Theory of the PI}
 
 The condition for parametric gain ${\cal R}$ when PI takes place in the Fabry-Perot cavity in the interferometer(excluding the impact of anti-Stokes
mode) has the following form\cite{bsv}:
\begin{eqnarray} 
\label{equ:gain} {\cal R}=\frac{\Lambda_1 W\omega_1}{cLm\omega_m\gamma_m\gamma_1}\times
\frac{1}{1+\frac{\Delta^2}{\gamma_1^2}}>1 \ , \\
\label{equ:ovlapfactor} \Lambda_1=\frac{V(\int A_0(r_{\bot})A_1(r_{\bot})u_z dr_{\bot})^2}{\int|A_0|^2
dr_{\bot}\int|A_1|^2dr_{\bot}\int|\vec{u}|^2 dV} \ ,
\end{eqnarray}
where $c$ is the speed of light, $L$ is the length of the Fabry-Perot cavity,  $m$ is the mirror's mass, $W$ is the power inside the cavity,
$\gamma_m$ and $\gamma_1$ are the relaxation rates of the elastic and Stokes modes respectively and $\Delta=\omega_0-\omega_1-\omega_m$ is the detuning value. The parameter $\Lambda_1$ is the overlapping factor between the elastic and the main  and Stokes optical mode, $A_0$ and $A_1$ are the optical field distributions at the  surface of the mirror for the optical main mode and the Stokes mode. The vector $\vec{u}$ is the spatial displacement of the elastic mode and $u_z$ is the z-component of $\vec{u}$  along the cylindrical axis. $\int d{r}_{\bot}$ are the integration over  reflecting surface of the end mirror and $\int dV$ -- over the mirror's volume $V$.

 \begin{table}[t]
  \centering
    \begin{tabular}{cccc} 
   $\omega_m/2\pi$, Hz & m & Stokes mode &$\cal{R}$ \\
   \hline
17341 & 2 &$LG_{02}$ &3\\
32103 &0 &$LG_{10} $&2.2\\
19579 &0 &$LG_{20} $&11.3\\
34508 &2 &$LG_{12} $&9.2\\
34626 &0 &$LG_{20} $&11.1\\
37855& 1 &$LG_{31}$& 1.2\\
38751& 0 &$LG_{40}$& 1.4\\
39309& 0 &$LG_{40}$& 1.7\\
 \hline
    \end{tabular}%
    \caption{The unstable elastic and Stokes optical
modes in the ET with sapphire mirrors. The elastic mode is
characterized by the azimuthal dependence $e^{im\varphi}$. Stokes optical modes have a Laguerre-Gauss $LG_{nm}$ profile with the radial index $n$ and  the
azimuthal index $m$. }
  \label{tab:addlabela}%
\end{table}%

 \begin{table}[t]
  \centering
    \begin{tabular}{cccc}
     $\omega_m/2\pi$, Hz & m & Stokes mode &$\cal{R}$ \\
   \hline
32207& 2 &$LG_{02}$ & 1.4\\
19456& 2 &$LG_{12}$& 1.5\\
34391& 2 &$LG_{12}$& 2\\
20858& 1 &$LG_{21}$ & 2.7\\
36749& 0 &$LG_{30}$ &2.9\\
38017& 3 &$LG_{23} $&1.7\\
38070 &3& $LG_{23}$ &1.8\\
39914 &0& $LG_{40} $&1.5\\
37325 &1&$ LG_{31} $& 2.3\\
35725& 0& $LG_{30}$& 1.2\\
35922 &1& $LG_{21}$ & 1.3\\
35428 &3& $LG_{13}$& 1.4\\
 \hline
    \end{tabular}%
    \caption{The unstable elastic and Stokes optical
modes in the ET with silicon mirrors. The elastic mode is
characterized by the azimuthal dependence $e^{im\varphi}$. Stokes optical modes have a Laguerre-Gauss $LG_{nm}$ profile with the radial index $n$ and  the
azimuthal index $m$.}
  \label{tab:addlabel1}%
\end{table}%

To predict the effect of PI it is necessary to know the magnitudes of the detuning and the overlapping factor(see formulas (\ref{equ:gain}) and (\ref{equ:ovlapfactor})).
As already mentioned, the frequencies and the optical field distributions  of the Gaussian optical modes is easy to calculate analytically, even in those cases
when the mirrors of Fabry-Perot cavities differ slightly from each other. In \cite{2008,2008_1,meleshko}  the methods of accurate calculation of  frequencies and
the displacement vectors for each elastic mode are discussed.

\section{PI in ET}

For a Fabry-Perot cavity of ET we use the parameters shown in the Table \ref{tab:addlabel}, where $\lambda$ is the wavelength of the main optical mode, $R_1, R_2$ are the radii of curvature of the input and end test masses of the Fabry-Perot cavity, $E$ -- Young's modulus, $\rho$ -- density, $\sigma$ -- Poisson's ratio of the mirror's material. The relaxation rates  of the elastic modes are calculated using loss angle $\phi=10^{-8}$. We assume that the relaxation rates of the Stokes optical modes  depend  both on the coefficient of energy transmission for the input test mass and diffractional losses of the mirrors of Fabry-Perot cavities.

We have estimated the number of unstable combinations of elastic and Stokes optical modes both for the case of silicon cavity mirrors  and the
mirrors made of sapphire. Elastic modes  frequencies and field of the deformations were evaluated by using the numerical package COMSOL$^\circledR$ in the frequency range up to 40kHz(the number of mesh elements  was around 40000). For the configuration of the ET we used the Fabry-Perot cavities  with a radius of the light beam waist $w_0\simeq 1.42$cm.  It is worth noting that all combinations of elastic and Stokes modes was deduced up to the optical modes of the ninth order taking into account the  diffraction losses of the Stokes modes.

In the Tables  \ref{tab:addlabela}  and \ref{tab:addlabel1}  all unstable combinations of the elastic and Stokes modes in the frequency range of the elastic modes and parametric gain values for mirrors made of sapphire and silicon were performed, respectively. The number of unstable modes in the ET
exceeds the number of unstable modes   for a LIGO interferometer (the mass of the mirrors in the interferometer LIGO is 40 kg, the mirrors were made of
fused silica). This statement can be explained  by the fact that the density of elastic modes in the spectrum is proportional to the third degree of the ratio of mirror size and the speed of sound in the material. This also leads to the fact that the number of unstable modes in the case of the mirrors made of silicon is more than for  the mirrors made of sapphire (see Tables \ref{tab:addlabela}  and \ref{tab:addlabel1}). Another important reason for large number of unstable modes is the increase in the optical mode density in the spectrum for the ET due to a 2.5-fold increase in the length of its Fabry–Perot cavities(arm cavity length 10000m in ET vs. 4000m in  Advanced LIGO).

\section{Results and Conclusion}

In this Letter we have analyzed two possible configurations of gravitational wave detectors of third generation like ET with the  mirrors of Fabry-Perot   cavities   made both  of silicon and sapphire and determined the number of unstable modes. At present,   existing gravitational wave detectors operate at room temperatures mainly with mirrors fabricated from fused silica which has a high value of mechanical  Q-factor($Q \simeq 2 \times 10^7$). However, in order to increase the sensitivity of the third-generation  detectors    larger values of Q-factors are  required, and the detectors must operate at the  temperatures substantially lower than room temperature. Unfortunately, the   quality factor Q of fused silica decreases with decreasing of temperature\cite{mat1,mat2}. At the same time,  silicon has good mechanical properties at low temperatures: high value of   quality factor(Q-factor reaches $10^9$), a small thermal expansion coefficient, etc. These facts make silicon a suitable material for use in third-generation gravitational wave detectors like ET.

However, silicon has a high absorption coefficient    at the wavelength 1064nm,which is equal to $\alpha \simeq 10 \rm{cm}^{-1}$(for fused silica $\alpha \simeq 2.5\times  10^{-7} \rm{cm}^{-1}$ )\cite{mat3}. In\cite{mat4} the authors proposed to use a silicon  with a laser at a wavelength of 1550nm, because silicon has a small optical absorption $\alpha \simeq   10^{-8} \rm{cm}^{-1}$ on this wavelength. Another important problem  for using silicon or sapphire is their anisotropic properties.
 
The results of this Letter  suggest that using mirrors made of silicon or sapphire will  increase the number of parametric instabilities in the Einstein Telescope  as compared to the LIGO interferometer. In  the frequency range of elastic modes up to 40kHz we have  found 8 unstable combinations
for sapphire mirrors and 12 combinations for silicon mirrors correspondingly.

It is  also useful to compare the number of unstable modes  both in ET  and in LIGO Voyager(blue design) for the mirrors made of silicon\cite{blair2017}. LIGO Voyager Blue design have practically the same parameters: laser wavelength is 2000nm, the mass of silicon mirrors with low acoustic loss is 204kg at the operating temperature of 123K and the arm cavity power 3MW. In \cite{blair2017} the parametric gains $\cal{R}$ for different elastic  modes have been calculated. At cavity power of 3MW in LIGO Voyager there are 1161 modes with ${\cal R}>0$  of which two modes have ${\cal R}>1$. The maximum parametric gain is 76.
Blair et al. have found the number of unstable elastic modes for different radii of curvature(ROC) of input test mass of Fabry-Perot cavity. The number of unstable modes varies from 2 to 5  for different ROCs.  It is easy to say that this result is in good agreement with the results calculated in this Letter. The number of unstable modes for ET with silicon test masses preliminary 2.5 times more than in  LIGO Voyager because the ratio of arm cavity lengths is $L_{ET}/L_{Voyager} = 2.5$, and the optical mode density increases preliminary in the same manner. 

In general, the same results were deduced in \cite{strigin2015} for  10000m ET arm cavity length with fused silica mirrors. The number of unstable modes varied from 5 to  9 for different ROCs. We have to say that our results have  only stochastical character because of variations in elastic parameters(Young's modulus, density and others). Therefore, the values of parametric gains can slightly fluctuate that influences the  number of possible PI unstable modes.  In this case the experimental results are required.

\section*{Acknowledgement}
The author kindly acknowledges the support of the Russian Foundation for Basic Research(grants nos. 11-02-00383-a and 14-02-00399-a).

\bibliographystyle{iopart-num}
\bibliography{refs}

\end{document}